%% file: main.tex
\documentclass[5p,times]{elsarticle}
\usepackage[numbers]{natbib}
\setcitestyle{square} 

\usepackage{url} 

\usepackage[T1]{fontenc}
\usepackage{amsmath, amsfonts, amssymb}
\usepackage[cal=boondox]{mathalfa} 
\usepackage{dsfont} 
\usepackage{relsize} 

\DeclareMathOperator*{\argmin}{arg\,min}
\newcommand{\N}{\mathbb{N}} 
\newcommand{\Z}{\mathbb{Z}} 
\newcommand{\Asterisk}{\mathlarger{*}}

\usepackage{booktabs} 
\usepackage{siunitx} 
\usepackage{tikz}
\usetikzlibrary{external} 
\usepackage{pgfplotstable}
\usepackage{pgfplots}
\pgfplotsset{compat=1.13}
\usetikzlibrary{pgfplots.groupplots}
\usetikzlibrary{lindenmayersystems} 

\usetikzlibrary{arrows, 3d} 
\usetikzlibrary{patterns}

\usepackage{subcaption}

\input{commands}

\newcommand{\revisionadd}[1]{#1} 

\begin{document}
\sloppy

\title{Scalable communication for high-order stencil computations using CUDA-aware MPI\tnoteref{t1}}
\tnotetext[t1] {
This work was supported by the Academy of Finland
ReSoLVE Centre of Excellence (grant number 307411); 
the European Research Council (ERC)
under the European Union's Horizon 2020 research and innovation
programme (Project UniSDyn, grant agreement n:o 818665); and
Theory within ASIAA from Academia Sinica.
}

\address[1]{Department of Computer Science, Aalto University, Konemiehentie 2, 02150 Espoo, Finland}
\address[2]{Academia Sinica, Institute of Astronomy and Astrophysics, Roosevelt Rd, 10617 Taipei, Taiwan}
\address[3]{Max Planck Institute for Solar System Research, Justus-von-Liebig-Weg 3, D-37077 G\"ottingen, Germany}
\address[4]{Nordita, KTH Royal Institute of Technology and Stockholm University, Roslagstullsbacken 23, SE-10691 Stockholm, Sweden}
\address[5]{Faculty of Science and Engineering, Åbo Akademi University, Tuomiokirkontori 3, 20500 Turku, Finland}

\author[1]{Johannes Pekkilä\corref{cor1}}
\cortext[cor1]{Corresponding author}
\author[2]{Miikka S. Väisälä}
\author[1,3,4]{Maarit J. Käpylä}
\author[1]{Matthias Rheinhardt}
\author[1,5]{Oskar Lappi}


\begin{abstract}
Modern compute nodes in high-performance computing provide a tremendous level of parallelism and processing power. However, as arithmetic performance has been observed to increase at a faster rate relative to memory and network bandwidths, optimizing data movement has become critical for achieving strong scaling in many communication-heavy applications. This performance gap has been further accentuated with the introduction of graphics processing units, which can provide by multiple factors higher throughput in data-parallel tasks than central processing units. In this work, we explore the computational aspects of iterative stencil loops and implement a generic communication scheme using CUDA-aware MPI, which we use to accelerate magnetohydrodynamics simulations based on high-order finite differences and third-order Runge-Kutta integration. We put particular focus on improving intra-node locality of workloads. 
\revisionadd{Our GPU implementation scales strongly from one to $64$ devices at $50\%$--$87\%$ of the expected efficiency based on a theoretical performance model. Compared with a multi-core CPU solver, our implementation exhibits $20$--$60\times$ speedup and $9$--$12\times$ improved energy efficiency in compute-bound benchmarks on $16$ nodes.}
\end{abstract}


\maketitle

\section{Introduction}
\label{sec:introduction}

Iterative stencil loops (ISLs) belong to a class of algorithms, in which data
points are updated by sampling their neighborhood in a fixed pattern called a stencil. ISLs,
or more generally, computations on a structured grid, have been
identified as one of the major recurring computational patterns in
high-performance computing (HPC) due to their prevalence in science and
engineering~\cite{asanovic2009}. Common applications include image
processing~\cite{mullapudi2015, kelley2014} and solving
partial differential
equations (PDEs)~\cite{brandenburg2003, keyes2013}. Because each data point can be
updated independently, ISLs can usually be processed efficiently on parallel computers.

Over the last ten years, compute nodes in HPC have been gradually shifting from homogeneous systems into systems housing multiple general-purpose processors and domain-specific accelerators; graphics processing units (GPUs) are the most commonly used ones. Of the TOP500 HPC systems, $27\%$ house one or more NVIDIA GPUs per node~\cite{top500}. As specialized co-processors, GPUs
can provide multiple times higher throughput in data-parallel tasks than central processing units (CPUs)\footnotemark, which makes them an attractive
platform for ISLs.
Optimization techniques for accelerating ISLs on a single GPU have been extensively studied in previous works~\cite{christen2011, datta2008}.

\footnotetext{A Tesla V100-SXM2-32GB GPU provides an operational performance
of $7.83$ TFLOPS (floating-point operations per second) and $863$ GiB/s
off-chip memory bandwidth~\cite{volta-whitepaper}, whereas a \revisionadd{20-core } Intel Xeon Gold
6230 CPU is capable of $1.25$ TFLOPS and supplying data at a rate of $131$
GiB/s~\cite{intel-whitepaper}.}

In computational sciences, large stencils
are often used to obtain sufficiently
accurate results. For example in astrophysical fluid simulations, the fluids
are frequently in a state of fully developed turbulence, and high-order
difference schemes, high-resolution discretization, and double-precision
arithmetic can be useful, or even mandatory, for discerning small-scale details.
In large-scale simulations, data movement is a likely bottleneck, as the amount of communication decreases at a lower rate than computation when the number of parallel processors is increased. This will be elaborated on in Section~\ref{sec:domain-decomposition}.

Reducing the performance impact of data movement is a notable
challenge. Wulf~\cite{wulf1995}, Patterson~\cite{patterson2004}, and
others~\cite{asanovic2006, hennessy2011}, observed that arithmetic performance
of microprocessors increases at a faster rate relative to the improvements
in memory bandwidth, and bandwidth improves at a faster rate than memory access latency. 
The performance of network interconnects has followed a similar trend. 
In a ten-year span, the operational performance of
a HPC node 
has
increased $26$-fold~\cite{top500}, whereas the
network interconnect bandwidth has increased only
$6.25$-fold\footnote{Infiniband QDR (2007) and HDR (2017)~\cite{mellanox-whitepaper}.}. As network bandwidth is generally an
order of magnitude less than off-chip memory bandwidth, optimizing
inter-node communication is critical for achieving efficient scaling
to a large number of compute nodes.

In this work, we address two major challenges with data movement in large-scale applications. Firstly, we estimate the upper bound for communication performance of $d$-dimensional stencil computations by defining a communication cost function for idealized hardware and solving an integer program to find the minimum required communication time. Secondly, we implement a scalable communication scheme, in which data movement latencies are hidden by pipelining computation and communication. We apply our implementation to a practical simulation setup commonly used in fluid dynamics research and compare the achieved performance to the theoretical maximum. The simulation setup employs high-order discretizations in space and time based on finite differences and Runge-Kutta integration methods.

The structure of this paper is as follows. In Section~\ref{sec:background},
we introduce the terminology used throughout this work to discuss the computational
aspects and scaling properties of ISLs. In Section~\ref{sec:performance-modeling},
we describe the performance model used for finding theoretical performance limiters
and evaluating the scaling of our implementation. In Sections~\ref{sec:domain-decomposition}
and~\ref{sec:implementation}, we present a communication cost function for stencil computations, find the upper bound for communication performance, and present the technical details
of our implementation. We give a brief description of the magnetohydrodynamics
solver used for benchmarks in Section~\ref{sec:magnetohydrodynamics}.
Finally, we present and discuss our results in Sections~\ref{sec:results} and~\ref{sec:discussion}, and conclude
the paper in Section~\ref{sec:conclusion}.

\section{Background}
\label{sec:background}

\begin{figure}
\centering
\input{tikz/fig-stencil}
\caption{Examples of two-dimensional stencils. The central cell (white) is updated by sampling the neighboring input points (gray). Here $r = 2$.}
\label{fig:stencil}
\end{figure}
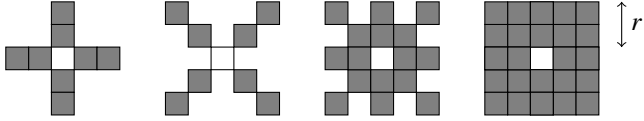

In ISLs, data points are updated by sampling their neighborhoods in a fixed memory access pattern, called a stencil (see Fig.~\ref{fig:stencil}).
The radius $r$ of a stencil is the maximal Chebyshev distance from $\boldsymbol{c}$
\begin{equation}
    r = \max_{\boldsymbol{s} \in \boldsymbol{S}} \left(\max_{i} |s_i - c_i| \right) \ ,
\end{equation}
where $\boldsymbol{c} = (c_0, c_1, \ldots, c_d)$ is the spatial index of the point being updated and $\boldsymbol{S}$ the set of stencil points. The exact shape of a stencil depends on the application. In the simplest case, a stencil contains all the points within its radius $\boldsymbol{S} = \{ \boldsymbol{s} \in \Z^d : |s_i| \leq r \}$. In this work, we focus on stencils of this form and its subsets with the same radius.

Data points are stored in a $d$-dimensional array, usually representing a structured grid with regular connectivity. In this context, we refer to data points as cells. During an iteration step, the cells belonging to the computational domain are updated according to some stencil operation. When updating cells near the boundaries, some stencil points fall outside
the computational domain. The set of these points is henceforth referred to as the \textit{halo}.
We use the term \textit{grid} to denote the set of all cells that belong to either the halo or the computational domain.

\begin{figure}
\centering
\input{tikz/fig-grid}
\caption{
Illustration of the computational domain partitioned into four subdomains. Boundaries of computational subdomains are marked with a solid line. A halo surrounds each subdomain, marked with a dotted line. One of the subgrids is highlighted with a dashed line for clarity.
}
\label{fig:grid}
\end{figure}
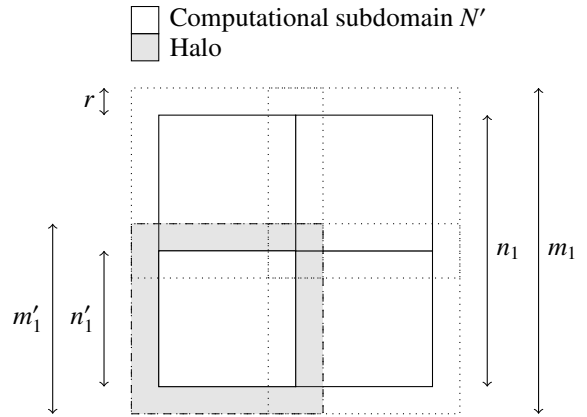

We use $d$-tuples of the form $\Phi = (\phi_1, \phi_2, \ldots, \phi_d)$ to denote domains, where $\phi_i \in \N$ is the number of cells in dimension $i$ and the total number of elements in $\Phi$ is
$C_\Phi = \prod_{i = 1}^d \phi_i$.
Using this notation, the domain of the grid is $M = (n_1 + 2r, n_2 + 2r, \ldots, n_d + 2r)$, where $N = (n_1, n_2, \ldots, n_d)$ is the computational 
domain (see Fig.~\ref{fig:grid}). When processing ISLs on distributed systems, $N$ must be decomposed into $C_P$
computational subdomains, with $p_i$ subdomains in dimension $i$.
Each computational subdomain $N'$ is also surrounded by a halo, forming a subgrid $M'$. For simplicity, we assume
all subdomains to have the same size and require, that each node is assigned exactly one subdomain. This enables us to regard $C_P$ henceforth as the number of nodes. 
The number of cells in $N'$ and $M'$ can now be written as 
\begin{equation}
C_{N'} = \prod_{i = 1}^{d} \tfrac{n_i}{p_i} \ ,~\text{and}
\end{equation}
\begin{equation}
C_{M'} = \prod_{i = 1}^{d} \left( \tfrac{n_i}{p_i} + 2r \right)\ .
\end{equation}
When processing ISLs on two or more nodes, a portion of the halo, local to one node, overlaps with the computational subdomain assigned to a neighboring node. After each update of the neighboring subdomain, the data corresponding to the halo segment must be communicated back to the initial node. This is called a halo exchange, as communication happens both ways for nodes sharing a boundary. 

\subsection{Previous work}

In previous work, we presented a library for accelerating ISLs on GPUs, called
Astaroth~\cite{pekkila2019}. It provides an application programming interface to the GPU resources and a domain-specific language \revisionadd{(DSL)} for writing stencil kernels. Astaroth was inspired by an earlier proof-of-concept
hydrodynamics solver presented in~\cite{pekkila2017, vaisala2017}, which was originally created for the purpose of exploring how to accelerate the computational methods used by the Pencil Code~\cite{pencilcode2020}.
Later, the library was extended to support computations on multiple devices on a single node
using CUDA peer-to-peer memory transfers~\cite{vaisala2020}. In this work, we extend Astaroth to
support computations on multiple nodes.
 
There are several libraries and frameworks designed for accelerating stencil codes. The one, which is closest related to Astaroth, is Physis~\cite{maruyama2011}, which has also been designed for accelerating stencil computations on GPUs and performs source-to-source translation from a DSL to CUDA and MPI. However, communication is carried out explicitly via host memory. Another library close to our work is LibGeoDecomp~\cite{schafer2008}, which is a mature, stencil-focused library supporting hierarchical geometric partitioning and load balancing on heterogeneous systems, including GPUs. Instead of a DSL, LibGeoDecomp provides C++ templates for describing the parameters for stencil kernels. Of PDE-specific libraries similar to our work, Fargo3D~\cite{benitez2016} is focused on accelerating MHD simulations, supports multiple GPUs and performs communication using CUDA-aware MPI. Instead of handling the memory of each GPU explicitly as in Astaroth, Fargo3D uses Unified Virtual Addressing (UVA) to manage the resources on a node. Yet another framework focused on advection-diffusion type problems is PyFR~\cite{witherden2014},
which provides hierarchical and graph-based partitioning based on the Metis~\cite{metis} and Scotch~\cite{scotch} software packages. The Cactus Framework has adopted
a more generic approach, providing tools for large-scale parallelization of various types of tasks, including stencil computations~\cite{goodale2003, jian2012}.

The main difference of Astaroth to existing libraries is its specialized focus on improving cache reuse in stencil computations, where the working set, that is, the data required to update a small group of cells, is too large to fit into the caches of a GPU. As such, Astaroth is especially suited for multiphysics simulations, which use high-order stencils, double
precision, and require data from several coupled fields to update a cell. For more details on the single-GPU optimization techniques and code generation of Astaroth, we refer the reader to~\cite{pekkila2019}.

\section{Methodology}
\label{sec:methodology}

\subsection{Performance modeling}
\label{sec:performance-modeling}

Performance models are useful for estimating theoretical performance limits, which can be used to determine whether further optimizations are needed or to calculate the expected scaling profile without having to queue for compute resources. In this section, we describe a simple performance model, which we use to find the upper bound for scaling performance. While the model has likely been introduced before, we are not aware of an established name.

In the following discussion, we use generic terminology and focus on ISL-specific definitions from Section~\ref{sec:domain-decomposition} onward. We use the term \textit{processing element} to refer to a generic computational unit that performs work in parallel, such as a node or a device. The terms \textit{host} and \textit{device} are used to refer to the CPU and GPU, respectively. \revisionadd{Throughout this work, we use the term \textit{CPU} to refer to the multi-core processor located on a single CPU socket.} Finally, we use the term \textit{communication} to refer to data movement within or between non-local memory systems.

As processing elements operate in parallel, the running time is the maximum time it takes for an element to complete its task. We denote the computational workload per processing element as $W$ data items and the amount of communication as $Q$ data items. Furthermore, $\pi$ is the operational capability of the hardware as data item updates per second, and $\beta$ the rate at which data elements can be communicated. The time taken by computation and communication is therefore $\tau_W = W \pi^{-1}$ and $\tau_Q = Q \beta^{-1}$, respectively. In this work, we measure $\pi$ empirically by benchmarking the program on a single device and calculate $\beta$ based on the theoretical network bandwidth and the size of a data item.

As computation and communication must be carried out in parallel to achieve efficient scaling, the running time of an ideal implementation is $\max \left( \tau_W, \tau_Q \right)$. Taking inspiration from Amdahl's law, we further include a term $\tau_0$ to capture the time taken in the sequential portion of the program. We use the term \textit{sequential} to refer to computations that cannot be carried out in parallel with communication.

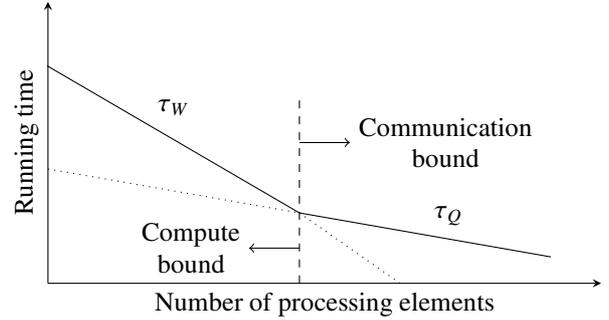
\begin{figure}
\centering
\input{tikz/fig-downhill}
\caption{
An example of the strong scaling profile produced with Eq.~\ref{eq:downhill}.
}
\label{fig:downhill}
\end{figure}

We can now model the running time as
\begin{equation}
\max \left( \tau_W, \tau_Q \right) + \tau_0 \ .
\label{eq:downhill}
\end{equation}
In this form, the model produces a scaling profile that is familiar from multi-processor benchmarks, see Fig.~\ref{fig:downhill}. 
When the performance of a kernel is limited by compute performance, it is said to be compute bound. In this case, $\tau_W > \tau_Q$. The kernel is communication bound when the opposite is true. Alternatively, we can express the performance bounds in terms of operational intensity $I=W/Q$, where the limiter is compute performance if $I > \pi/\beta$~\cite{williams2009}.

\subsection{Domain decomposition}
\label{sec:domain-decomposition}

There are two major considerations for implementing a communication scheme for distributed applications. Firstly, the problem domain must be decomposed into $P$ subdomains, and secondly, the subdomains must be assigned to processing elements. In this section, we use hierarchical geometric partitioning~\cite{teresco2006, zheng2011} to find a decomposition and processor assignment for ISLs that minimize the communication surface area. In this approach, the partitioning is optimized recursively on each level of the processing element hierarchy. We consider two levels: node- and device-level. Furthermore, we assume that the network topology is a fat tree and the devices within a node are fully connected. These assumptions imply that the bandwidth between any pair of nodes, or devices within a node, is roughly the same and it is possible to form parallel connections to arbitrary many neighbors.

Ultimately, the goal is to balance workloads across processing elements and minimize data
movement~\cite{niedermeier1997, zumbusch2000}.
If we consider Eq.~\ref{eq:downhill} to model running time at sufficient accuracy, we can minimize the communication surface area of the critical path by solving the integer programming problem
\begin{align}
    \begin{split}
        &\argmin_{P} \max \left( \tau_W, \tau_Q \right) + \tau_0\\
        &\text{subject to } p_i \in \N, \prod_{i = 1}^d p_i = C_P \ ,
    \end{split}
\label{eq:decomposition-ip}
\end{align}
where $\max \left( \tau_W, \tau_Q \right) + \tau_0$ is the worst-case running time for processing a subdomain. Throughout this work, we use the term \textit{optimal decomposition} to refer to $P$ which solves Eq.~\ref{eq:decomposition-ip}. First, we assume for the sequential portion of the program $\tau_0 = 0$. This implies that all computation and communication can be carried out in parallel, which is approximately the case when the implementation is sufficiently pipelined.

Next, we define $W$ and $Q$ for ISLs. In the following proofs, we first find the optimal node-level decomposition, and later expand the reasoning to include heterogeneous nodes containing multiple devices. The amount of local work per update step for each node is
\begin{equation}
W = C_{N'} = \frac{C_N}{C_P}\ .
\end{equation}
As $W$ does not depend on the choice of the components of either $N$ or $P$, we can focus on the case 
$\tau_W < \tau_Q$.
Furthermore, as
$\beta$
is a constant, the objective function in Eq.~\ref{eq:decomposition-ip} simplifies to $Q$.
To simplify the definition of $Q$, we assume that $p_i \geq 3,~\forall p_i \in P$.
If the boundaries are periodic, which is the case in our tests, then the following definition also holds when $p_i \geq 2,~\forall p_i \in P$. The number of cells communicated during a halo exchange per node in the worst case is
\begin{equation}
Q = 2 \left( C_{M'} - C_{N'} \right) \ .
\label{eq:qnp}
\end{equation}
The worst case behaviour is witnessed when all halo cells must be exchanged. This occurs when each of the boundaries of a computational subdomain faces another subdomain.

As $W$ is inversely proportional to $C_P$ with coefficient $C_N$, ISLs are expected to exhibit ideal scaling when the application is compute bound, that is, $\tau_W \geq \tau_Q$. We use the term \textit{ideal scaling} to refer to the case, where the performance grows linearly with the number of processing elements at $100\%$ efficiency. As $Q$ scales at a slower rate 
in comparison to $W$, scaling efficiency is reduced when the performance is bound by data movement.

For relatively low $C_P$, it is feasible to conduct an exhaustive search for the optimal decomposition. In other cases, a more sophisticated approach, such as using heuristics to reduce the search space, is likely needed. When using $P$ as a static mapping, the solution can be stored in a lookup table for quick access. The optimal components of $P$ for typical choices of $N$ are listed in Appendix A. On node level, the workloads are inherently balanced, as we require that $N'$ is the same for all nodes and the bandwidth between any pair is the same.

Next, we consider the case when a node contains multiple devices. The optimal decomposition can be found by recursively subdividing $N'$ further to $C_{P'}$ devices available on a node. We denote the sizes of the per-device grid and computational domain as $C_{M''}$ and $C_{N''}$, respectively. Similar to the node-level definitions,
\begin{equation}
C_{N''} = \prod_{i = 1}^{d} \tfrac{n_{i}'}{p_{i}'} \ ,~\text{and}
\end{equation}
\begin{equation}
C_{M''} = \prod_{i = 1}^{d} \left( \tfrac{n_{i}'}{p_{i}'} + 2r \right) \ .
\end{equation}
For the rest of this section, we refer to $N'$ as the computational domain and $N''$ as the computational subdomain.

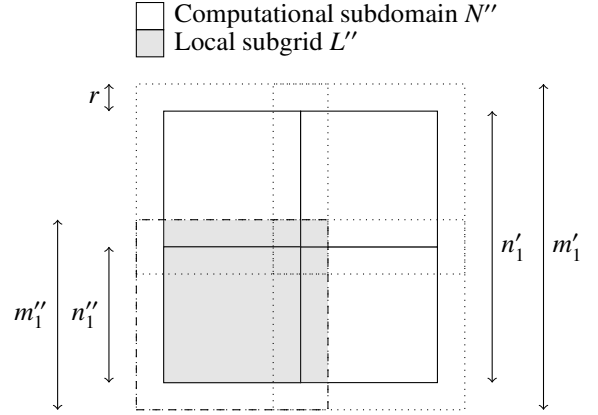
\begin{figure}
\centering
\input{tikz/fig-subgrid}
\caption{
Illustration of an intra-node computational domain decomposed into four subdomains. Boundaries of computational subdomains are marked with a solid line. A halo surrounds each subdomain, marked with a dotted line. One of the subgrids is highlighted with a dashed line for clarity. The local subgrid consists of the cells that are available without inter-node communication.
}
\label{fig:subgrid}
\end{figure}

To define the amount of communication performed via the intra-node communication fabric, we introduce the concept of a local subgrid $L''$. The local subgrid comprises the cells in the computational subdomain and the portion of the halo, that overlaps with the computational subdomain of any of the intra-node neighbors (see Fig.~\ref{fig:subgrid}). The size of the local subgrid is
\begin{equation}
C_{L''} \geq \prod_{i = 1}^{d} \left( \tfrac{n_i'}{p_i'} + r \cdot \mathds{1}_{p_i' \geq 2} \right) \ ,
\end{equation}
where $\mathds{1}_{p_i' \geq 2}$ is an indicator function. Intuitively, if $p_i' \geq 2$, then there is at least one intra-node boundary. 
As intra- and inter-node communication can be carried out in parallel, the total communication time per device is ideally
\begin{equation}
\max \left( \beta_{\text{inter}}^{-1} Q_{\text{inter}},~\beta_{\text{intra}}^{-1} Q_{\text{intra}} \right) \ .
\label{eq:qmin-inter-intra}
\end{equation}
Furthermore,
\begin{equation}
\begin{split}
    Q_{\text{inter}} &= C_{M''} - C_{L''}\\
    Q_{\text{intra}} &= C_{L''} - C_{N''} \ .
\end{split}
\end{equation}
The dominant factor in Eq.~\ref{eq:qmin-inter-intra} is typically $\beta_{\text{inter}}^{-1}Q_{\textit{inter}}$, as $\beta_{\text{inter}}^{-1} > \beta_{\text{intra}}^{-1}$ almost always on modern systems and
because
\begin{equation}
\begin{split}
    C_{M''} + C_{N''} &> 2 C_{L''} 
\end{split}
\end{equation}
holds for $N, P \in \N^d$ and $r \in \N$, we have
$Q_{\text{inter}} > Q_{\text{intra}}$.
Therefore we can find the optimal intra-node decomposition by minimizing $C_{M''} - C_{L''}$. By defining $C_{L''}$ as the lower bound for intra-node communication, minimizing $C_{M''} - C_{L''}$ maximizes worst-case performance. The optimal intra-node decompositions for $P'$ are listed in Appendix A.

We can compare the rate of decrease in communication of spatial decomposition schemes by varying the degrees of freedom of $P$. For example in a one-dimensional decomposition, regardless of the dimensionality of the grid, only one component of $P$ is free while the others are bound to unity. The scaling of data movement in common spatial decomposition schemes is illustrated in Fig.~\ref{fig:decomposition-scaling}. While a one-dimensional decomposition is easy to implement and
scales reasonably well to a low number of nodes~\cite{vaisala2020}, it is clear that multi-dimensional decomposition is required for large-scale applications.
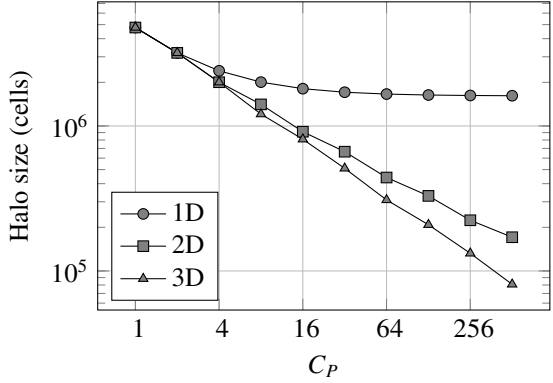
\fig{decomposition-scaling}{The size of the halo as a function of $C_P$ in one-, two-, and three-dimensional decomposition schemes.}{ylabel = Halo size (cells), xlabel = $C_P$, , legend pos = south west}

\subsection{Implementation}
\label{sec:implementation}

\begin{figure}
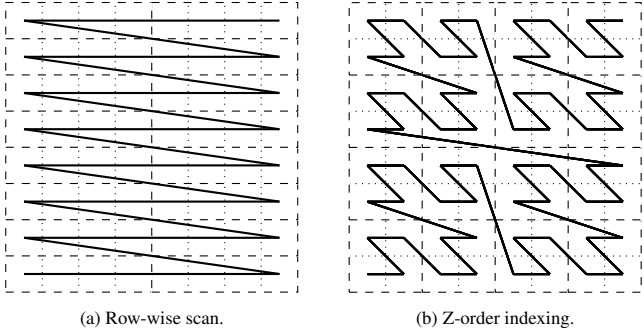

    \centering
    \begin{subfigure}{0.49\columnwidth}
        \centering
        \rowwisescan
        \caption{Row-wise scan.}
        \label{fig:row-wise-scan}
    \end{subfigure}\hfill
    \begin{subfigure}{0.49\columnwidth}
        \centering
        \mortonorder
        \caption{Z-order indexing.}
        \label{fig:morton}
    \end{subfigure}
    \caption{
The mapping of subdomains to devices using row-wise scan and two-dimensional Z-order indexing. Device and node boundaries are indicated with dotted and dashed lines, respectively.
}
\label{fig:sfc}
\end{figure}

In our implementation, we subdivide the computational domain recursively along each axis in succession and use Z-order indexing~\cite{morton1966} to map processors to subdomains (Fig.~\ref{fig:morton}). The partitoning is given by $P^{\Asterisk} = \text{morton}^{-1}(C_{{P}^{\Asterisk}} - 1) + (1, 1, 1)$, where $C_{P^{\Asterisk}} = C_{P} C_{P'}$. The function $\text{morton}(\varphi) = i$ interleaves the binary representation of a multidimensional coordinate $\varphi$ to obtain a one-dimensional index $i$, and $\text{morton}^{-1}(i) = \varphi$ is its inverse operation. For example, $\text{morton}^{-1}$ maps a binary index $i = \texttt{abcdef}_{2}$ to coordinate $\varphi = (\texttt{cf}_{2}, \texttt{be}_{2}, \texttt{ad}_{2})$.

\revisionadd{The Z-order curve preserves locality to a relatively high degree, meaning that
one-dimensional indices along the curve are likely mapped to multidimensional
coordinates that are spatially nearby. If processes within a node
are assigned contiguous MPI ranks, Z-order indexing can be used to enhance
intra-node locality of the subdomains.}
By comparison with the communication-optimal decomposition discussed in Section~\ref{sec:domain-decomposition} and Appendix A, the Z-order mapping minimizes, or nearly minimizes, the data movement on the critical path in the case where $n_x = n_y = n_z$ and $C_{P'} = 4$. 

In contrast to more intuitive processor mappings, locality-preserving space-filling curves can provide better load balancing and reduced data movement. Consider the case where subdomains are assigned to processors in a row-wise scan pattern (Fig.~\ref{fig:row-wise-scan}). In this case it is possible for neighboring processors to communicate a different amount of data to inter-node neighbors, which incurs a load imbalance. In the three-dimensional case, $C_{P'} = 8$, $C_{P} >> C_{P'}$, and the dimensions of subdomains are equivalent, the number of faces shared with inter-node neighbors ranges from four to five. With Z-order indexing, each process has exactly three inter-node-facing edges. The amount of data communicated along the critical path is also larger with row-wise indexing, as the worst-case size of the halo segments is larger when decomposing the intra-node domain in one dimension instead of three.

The use of space-filling curves in large-scale computations has been explored \revisionadd{by, for example, Tsuzuki~\cite{tsuzuki2016} and Li~\cite{li2018}}.
It should 
be noted,
however,
that on practical hardware, pairing our communication cost function with an established graph-based partitioner, such as Scotch~\cite{scotch}, may yield 
even
higher-quality decompositions. In this work, we determine that Z-order \revisionadd{mapping} is sufficiently communication-efficient for our purposes, and leave rigorous comparisons with more established methods for future work.

Several MPI implementations, notably MVAPICH and OpenMPI, provide support for transfers to/from CUDA-allocated device
memory. As memory transfers are routed automatically via the fastest communication fabric and the programmer can treat device pointers simply like host pointers, implementing device-to-device communication with CUDA-aware MPI is straightforward. However, special care is needed to ensure the correct pipelining of compute kernels and data transfers.

Furthermore, the GPUDirect remote direct memory access (RDMA) technology has been introduced to enable low-latency inter-node device-to-device communication directly via the network interface controller, bypassing host memory. However, due to lower bandwidth, device-to-device RDMA generally provides better performance only when sending small messages of size $32$ KiB or less~\cite{potluri2013}. Larger messages are buffered through host memory in a pipelined fashion. Almost all messages sent in our implementation are above this threshold.

\input{tikz/fig-sfc}

Executing memory and compute operations in parallel is necessary to hide communication latencies. On a single device, the CUDA API provides concurrency primitives, called streams, for managing the asynchronous execution of kernels. On the multi-node level, concurrency can be managed using the non-blocking variants of the send and receive functions provided by MPI.

\begin{figure}
\centering
\input{tikz/fig-segments}
\caption{Halo segments. Each segment is sent to a unique neighbor. The inner and outer segments of the computational subdomain are not visible.
}
\label{fig:halo-segments}
\end{figure}
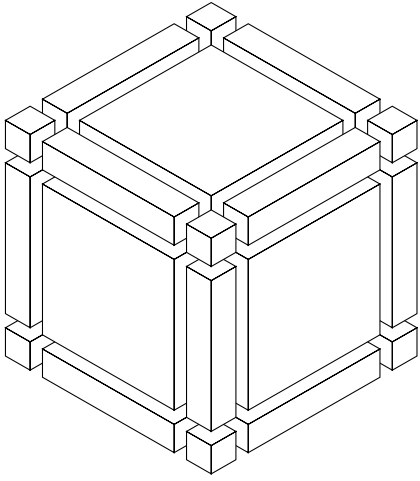

To carry out computation in parallel with communication, we divide the computational domain conceptually into one inner and several outer segments. The inner segment can be updated without information from the halo, while the outer segments can be updated only after communication has finished.
In three-dimensional grid decompositions, the domain of the inner segment comprises $(n_{x}' - 2r, n_{y}' - 2r, n_{z}' - 2r)$ cells. Similar to the inner and outer segments, we partition the halo into conceptual segments, where each segment overlaps with the computational domain of a distinct neighbor. The halo segments are illustrated in Fig.~\ref{fig:halo-segments}. From largest to smallest, we call the segments sides, edges, and corners. There are 6 side segments, 12 edge segments, and 8 corner segments. Each halo segment can be uniquely identified by the spatial index of its first element, which we use as a message tag. A segment at index $s_i$ is mapped to the computational domain of the receiving device as $s_i' = \left( \left( s_i - r \right) \mod n_i' \right) + r$.

Before initiating halo exchange, the data elements corresponding to each segment are packed into a contiguous buffer. 
This has two advantages. Firstly, the bulk of memory operations is performed within the faster local memory and secondly, the throughput for sending a few larger messages is generally higher than sending several smaller ones. Two buffers are allocated corresponding to each segment in order to send and receive in parallel. The first buffer is used for packing and sending, and the second for receiving and unpacking. Each ISL iteration consists of the following steps.
\begin{itemize}
\item \textbf{Inner segment update}.
Update cells in the inner segment.
\item \textbf{Packing}.
Pack outbound halo segments into contiguous buffers.
\item \textbf{Halo exchange}.
Exchange halo segments with neighboring devices.
\item \textbf{Unpacking}.
Unpack inbound halo segments into the local halo.
\item \textbf{Outer segment update}.
Update cells in the outer segments.
\end{itemize}

To execute packing, halo exchange, and unpacking in a pipelined fashion, each segment is associated with a distinct non-blocking CUDA stream. To exchange the segments, we use functions \verb!MPI_Isend! and \verb!MPI_Irecv!. The update of the inner segment is launched simultaneously with the halo exchange. Because the halo exchange depends on packed data being available, it is critical to ensure that packing completes before starting the inner segment update. This enables concurrency of computation and communication. We handle this by assigning higher priorities to packing streams, but one could alternatively add an additional synchronization step after packing.

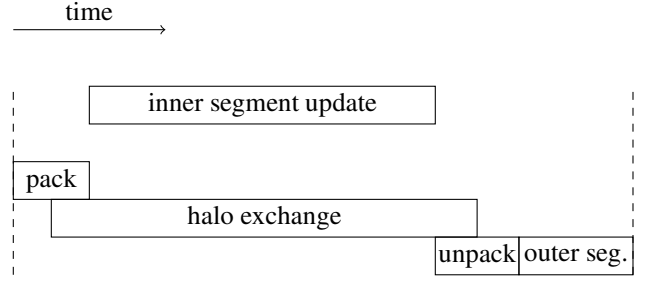
\begin{figure}
    \centering
    \input{tikz/fig-iteration}
    \caption{Functions applied during a single iteration. Synchronization is indicated with a dashed line. The time taken by each function is not to scale for illustrative purposes.}
    \label{fig:concurrency}
\end{figure}

After all data segments have been received, as indicated by the completion of \verb!MPI_Waitall!, we launch the kernels for updating the outer segments. The dependencies and execution order of these tasks are illustrated in Fig.~\ref{fig:concurrency}. After each iteration, the devices are synchronized using \verb!cudaDeviceSynchronize! and \verb!MPI_Barrier!. The implementation supports both single and double precision.

\section{Implementation example: Magnetohydrodynamics}
\label{sec:magnetohydrodynamics}

A particularly 
active domain of application for Astaroth is in astrophysical fluid dynamics, especially
the study of magnetized astrophysical plasma dynamics
in the magnetohydrodynamics (MHD) framework (for a general introduction of astrophysical MHD see e.g. \cite{Shu1992PAVII}).
MHD is based on the approximation, that if charged plasma particles are highly collisional, resistivity is low, and the explored length and time scales are much larger than the ion gyroradius and their oscillation times, plasma can be modelled as a conducting fluid coupled with a magnetic field. 

Astrophysical MHD problems are usually highly non-linear and require substantial computing resources because high resolutions in time and/or space,
and long integration times, due to the vastly differing time scales of turbulence and the phenomena of interest,
are required in realistic setups. In addition, problem sizes inflate when an extended parameter space has to be scanned. 
MHD has a wide range of applications in multiple astrophysical domains. It is used to study phenomena such as solar magnetic activity, the Earth's magnetosphere, interstellar medium, and star formation. The same methods can be also used in general fluid mechanics, 
because when neglecting the magnetic field, the
MHD equations reduce to the standard equations of hydrodynamics.

The self-contained MHD code has been utilized in a recent work~\cite{vaisala2020}, with single-node parallelization, to explore the 
kinematic growth phase
of turbulent MHD dynamos
in the isothermal regime. 
Our MHD solver follows the approach of the \textit{Pencil Code}: we use a non-conservative but high-order finite difference method to explore the non-linear problems. 
Our implementation supports $2$nd-, $4$th-, $6$th-, and $8$th-order central finite differences and time integration by a third-order Runge-Kutta (RK) method~\cite{williamson1980, brandenburg2003}. 
The stencils used to compute the derivatives contain the points
\begin{equation}
    \boldsymbol{S} = \bigl\{ z \left(\boldsymbol{x} \pm \boldsymbol{y} \right) : 
        \left\{\boldsymbol{x}, \boldsymbol{y} \right\} \subset \left\{ \boldsymbol{i}, \boldsymbol{j}, \boldsymbol{k}, \boldsymbol{0} \right\}, |z| \leq r, z \in \Z \bigr\} \ ,
\label{eq:stencil-fd}
\end{equation}
where $\boldsymbol{i},\boldsymbol{j},\boldsymbol{k}$ is the standard basis of a 3D space. We use the term $k$th-order stencil to refer to stencils used to compute $k$th-order accurate central finite differences, where $k = 2r$. The related stencils are illustrated in Fig.~\ref{eq:stencil-fd}. The full set of MHD equation is listed in Appendix B. For more details, we refer the reader to~\cite{pekkila2019} and~\cite{vaisala2020}.
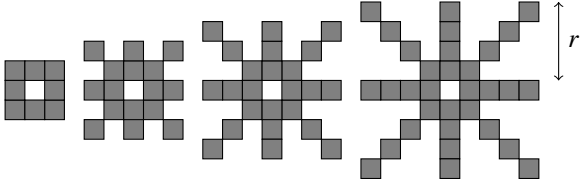
\begin{figure}
\centering
\input{tikz/fig-stencil-fd}
\caption{Two-dimensional cuts of the stencils used to simulate magnetohydrodynamics in this work. Illustrated from left to right: second-, fourth-, sixth-, and eight-order stencils. See Eq.~\ref{eq:stencil-fd} for the definition of the three-dimensional shape.}
\label{fig:stencil-fd}
\end{figure}

\section{Results}
\label{sec:results}

The tests were conducted on a cluster consisting of a total of 80 SuperServer 1029GQ-TVRT nodes connected in a fat tree network~\cite{superserver-manual}. Each node houses two Intel Xeon Gold 6230 Cascade Lake $20$-core processors running at $2.1$ GHz and four Tesla V100-SXM2-32GB GV100GL (rev a1) GPUs running at $1.53$ GHz. 
\revisionadd{The stated thermal design power (TDP) of an Intel Xeon Gold 6230 CPU is $125$ watts~\cite{intel-whitepaper}, and the stated TDP of a GV100GL GPU is $300$ watts~\cite{volta-whitepaper}.} 
Each GV100GL is connected to the other three GPUs via pairs of NVLink 2.0 connections, providing $91$ GiB/s bidirectional bandwidth per pair. The total intra-node NVLink bandwidth per GPU is $270$ GiB/s. Each node houses two Mellanox ConnectX-6 InfiniBand HDR100 MT28908 adapters, which provide $23$ GiB/s bidirectional inter-node bandwidth per adapter~\cite{mellanox-whitepaper}. Error-correcting codes (ECC) were enabled in all tests. We confirmed the transfer rates by measuring the time to transfer $12$ MiB data blocks, which is the same size as the largest individual halo segment transferred in simulations employing $256^3$ cells. The effective device-to-device intra-node bandwidth was $86$ GiB/s and inter-node bandwidth $40.8$ GiB/s.


Astaroth~\cite{astaroth-repository}, commit \verb!3804e72!, was compiled using GCC $8.3.0$, CUDA toolkit $10.1.168$, and OpenMPI $4.0.3$. One MPI task was assigned per GPU. Each \revisionadd{multi-core }CPU controlled two GPUs closest to it in the node topology and a pair of GPUs shared access to one network interface controller. The rendezvous protocol used by the Unified Communication X (UCX) framework was configured by setting \verb!UCX_RNDV_THRESH=16384!, \verb!UCX_RNDV_SCHEME=get_zcopy!, and \verb!UCX_MAX_RNDV_RAILS=1!, as this gave the best performance on the tested hardware. It should be noted, that the optimal configuration for the rendezvous protocol is system specific. 

\revisionadd{To evaluate whether our implementation is competitive with established work used in production,} we compared the scaling performance of Astaroth with that of Pencil Code~\cite{pencilcode2020}, commit \texttt{7ddde40}. The simulation setup is available at~\cite{pencil-code-test-setup}.

\revisionadd{The Pencil Code (PC) was benchmarked on two clusters: Puhti and Mahti. The Puhti benchmarks were run on a CPU-only partition of the same cluster as the GPU tests, which also houses two Intel Xeon Gold 6230 processors but only one Mellanox HDR100 network interface controller (NIC) per node. The effective bidirectional inter-node bandwidth in our experiments was roughly $20$ GiB/s.
The network topology of the Puhti system is a fat tree~\cite{leiserson1985}. A compute node on the Mahti system houses two AMD Rome 7H12 $64$-core CPUs running at $2.6$ GHz, where each multi-core CPU is split into $4$ NUMA domains and the TDP of a single CPU is $280$ watts~\cite{amd-whitepaper}, $256$ GiB memory, and a single Mellanox HDR200 NIC. The network topology on Mahti is Dragonfly+~\cite{shpiner2017}. The effective bidirectional inter-node bandwidth was roughly $39$ GiB/s. We compiled PC using the highest optimization level (\texttt{O2}) recommended for production runs and used compilers tuned for the clusters. On Puhti, we used Intel compiler version $19.0.4$ and HPCX-MPI $2.4.0$, and on Mahti, Intel compiler $19.1.1$ and OpenMPI $4.0.3$. We refer to PC benchmarks run on the respective systems as PC-Puhti and PC-Mahti hereafter.}

Pencil Code was chosen for comparison due to the following reasons: First, it is a mature MHD-solver widely used for production in large-scale astrophysical simulations~\cite{pencilcode2020-scientific-usage}. Secondly, Astaroth and Pencil Code use the exact same methods and equations in their MHD solvers. Thirdly, comparing CPU and GPU scaling gives us an indicator, whether GPU-applications can be competitive with established CPU solvers in large-scale HPC applications \revisionadd{in terms of throughput and energy efficiency}. The perceived bottleneck of CPU-GPU communication has been a cause of concern. Fourthly, Pencil Code uses simple axis-wise partitioning supplied by the user and row-wise scan indexing to map subdomains to processors, in contrast to Astaroth's Z-order mapping that favors the assignment of intra-node neighbors to nearby subdomains. Finally, CPU and GPU applications both use the same communication fabric and implementing a communication scheme for either architecture is code-wise nearly identical. While computation throughput is higher on GPUs in our test case, communication-wise there is little difference with CPU applications.

In the benchmarks, we measured the time to complete a third-order Runge-Kutta integration step. The simulation variables were initialized to random values in the range $[0, 1]$ and the timestep was set to a constant $\delta t = 1.19209 \times 10^{-7}$. The simulation was run for $100$ warm-up steps before measuring the running time of $1000$ integration steps. As an exception, we timed only $100$ steps in tests which included $\geq 512^3$ cells due to longer integration times. Double precision was used in all tests.

We verified the results by comparing the simulation state after an integration step with a model solution, which was obtained with a single-core CPU solver logically equivalent with the GPU solver. For a model value $\mathcal{m}$, the unit in the last place (ulp) is
\begin{equation}
\epsilon = 2^{\lfloor \log_{2}| \mathcal{m} | \rfloor - \mathcal{p} + 1} \ ,
\end{equation}
where $\mathcal{p}$ is the precision of the floating-point number. For double-precision, $\mathcal{p} = 53$. The absolute error in ulps for a candidate value $\mathcal{c}$ is calculated as
\begin{equation}
\frac{|\mathcal{m} - \mathcal{c}|}{\epsilon} \ .
\end{equation}
We verified the solver on $1$--$16$ devices in problem sets consisting of $256^3$ cells. In all cases, the maximum absolute error was $\le 2$ ulps. We deemed this to be within acceptable limits, as slight round-off errors are expected to accumulate when rounded intermediate values are used in calculations.

We tested our implementation in \revisionadd{six} benchmarks. In all of the following figures, the performance is shown as the median time per cell per integration step, where each step comprises three ISL iterations. Sixth-order stencils were used unless otherwise mentioned. \revisionadd{We confirmed that processes within a node were assigned contiguous MPI ranks as assumed in Section~\ref{sec:implementation}.}

Firstly, we measured the effective integration time per cell for various grid sizes (Fig.~\ref{fig:results-meshsize}). For problem sizes consisting of $256^3$, $512^3$, and $1024^3$ cells, we saw scaling to $64$ devices at $18\%$, $43\%$, and $87\%$ parallel efficiency, respectively, compared to ideal scaling. Theoretical strong scaling, calculated using the performance model discussed in Section~\ref{sec:performance-modeling}, is shown in Fig.~\ref{fig:results-model}. We used $\pi^{-1} = 2.2$ ns as the computational performance, which we determined empirically by \revisionadd{measuring} the integration step time on a single device. The communication performance $\beta^{-1} = 3.9$ ns was determined theoretically based on the cell size in bytes and the maximum network bandwidth of $46$ GiB/s. 
Compared with the theoretical model, the effective performance was $50\%$, $59\%$, and $87\%$ of the theoretical maximum on $64$ devices. Our implementation exhibited near-ideal scaling when compute-bound, whereas there was a notable drop in scaling efficiency when communication started to dominate.
\begin{figure}
    \centering
    \begin{minipage}[t]{0.49\textwidth}
        \centering
        \figg{results-meshsize}{ylabel = Time per cell (ns), xlabel = Devices, log ticks with fixed point}
        \caption{Effective strong scaling.}
        \label{fig:results-meshsize}
    \end{minipage}\hfill
    \begin{minipage}[t]{0.49\textwidth}
        \centering
        \figg{results-model}{ylabel = Time per cell (ns), xlabel = Devices, log ticks with fixed point}
        \caption{Theoretical strong scaling.}
        \label{fig:results-model}
    \end{minipage}
\end{figure}


In our second test, we isolated computation and communication to study the performance of the system further. Our results are presented in Fig.~\ref{fig:results-isolated}. As can be seen, 
communication
on a single node
(up to four devices)
is completely hidden, which is enabled by the relatively high device-to-device bandwidth and the devices being fully connected within a node. Inter-node communication is more sensitive to increases in 
the amount of communicated data.
The portion of the halo communicated to inter-node neighbors increases gradually from 1 to 8 nodes (4 to 32 devices). This can be seen as a stagnation in communication times between 4 to 16 devices. The effect from 16 to 32 devices is significantly weaker and no longer visible. 
Ideally, the measured integration time is the maximum of computation and communication times. This is not the case, and we can see an overhead of roughly $15$--$20$\% compared to communication and communication benchmarked individually. The overhead is \revisionadd{likely} caused by the sequential portion of our implementation. We also measured the overhead caused by synchronization, which was $0.02$--$2\%$ of the integration time. The stencils used in this work did not require the communication of the corner halo segments, but this is not the case for all applications. For completeness, the benchmark employing the communication of the corner segments is also shown. The corner segments were small enough to trigger RDMA. The added communication of the corner halo segments increased running times roughly by $0.03$--$10\%$.
\fig{results-isolated}{
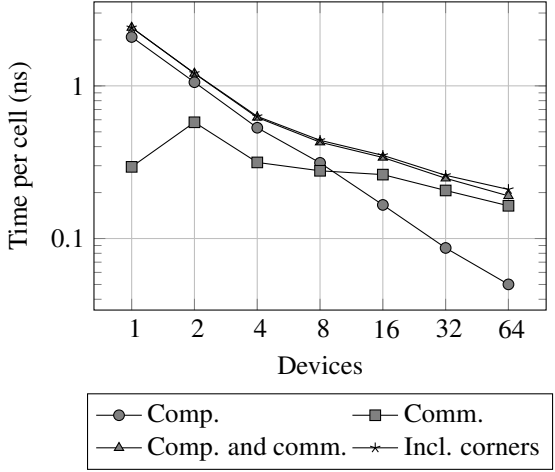
Compute and communication times measured separately in a $256^3$-cell simulation. The overhead caused by synchronization is negligible and has been left out for clarity.
}{ylabel = Time per cell (ns), xlabel = Devices, log ticks with fixed point, legend columns = 2, legend style={at={(0.5,-0.27)},anchor=north}}

\begin{figure}
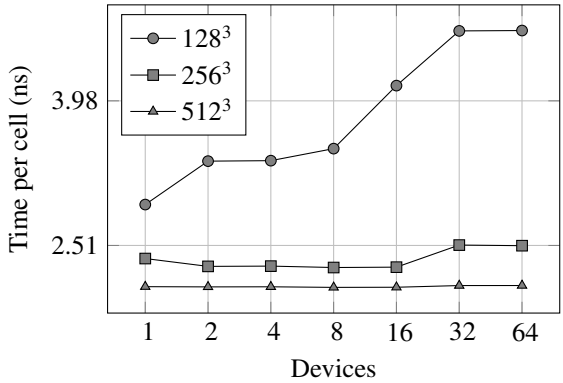

    \centering
    \figg{results-weak}{ylabel = Time per cell (ns), xlabel = Devices, log ticks with fixed point, legend pos = north west}
        \caption{Weak scaling.}
        \label{fig:results-weak}
\end{figure}

In our third test, we measured weak scaling by assigning each device the same amount of work (Fig.~\ref{fig:results-weak}). The interconnect boundaries are clearly visible in the figure. The onset of intra-node communication can be seen as the transition between one and two devices, and the onset of inter-node communication between four and eight. The proportion of the node-level halo communicated to inter-node neighbors increases gradually from four devices, reaching the maximum at $32$ devices ($8$ nodes).

\begin{figure}
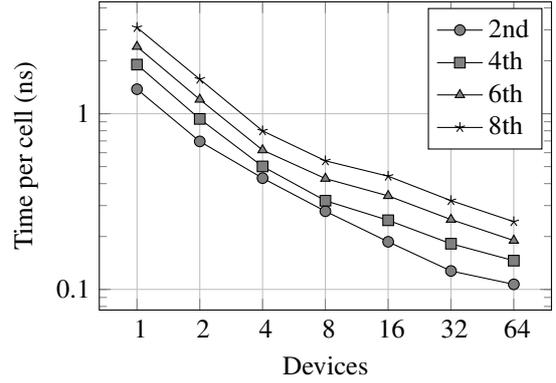

    \centering
    \figg{results-order}{ylabel = Time per cell (ns), xlabel = Devices, log ticks with fixed point}
        \caption{Strong scaling of varying stencil orders.}
        \label{fig:results-order}
\end{figure}

In our fourth test, we measured scaling of varying stencil orders in a $256^3$-cell simulation (Fig.~\ref{fig:results-order}). As expected, low-order stencil computations
require less communication and scale more efficiently than higher-order ones. Second-order stencils scale relatively efficiently to $32$ devices, whereas $8$th-order stencils computations become bandwidth-bound at $8$ devices.

\begin{figure*}
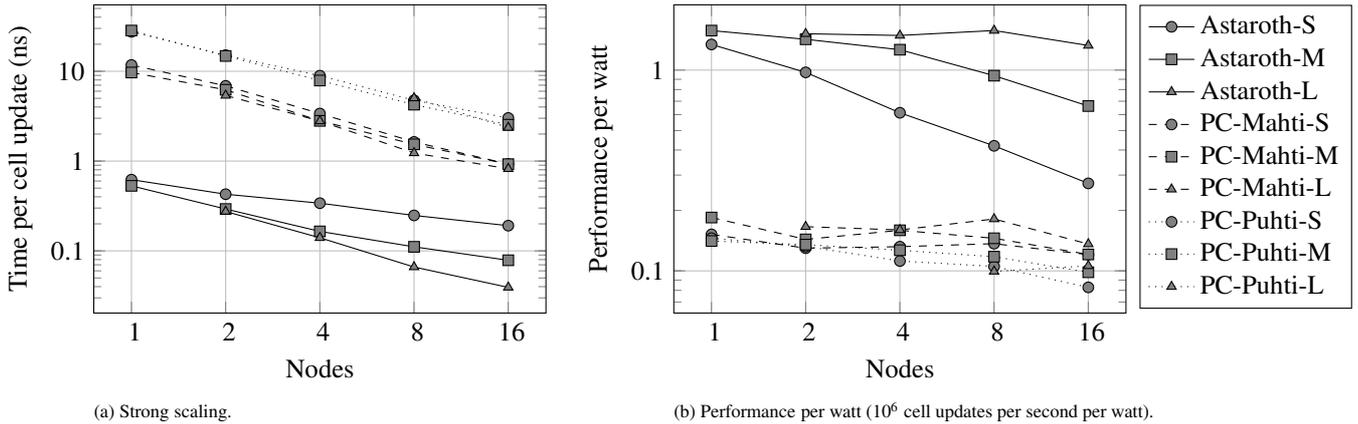

        \centering
        \subfloat[Strong scaling.]{
        \label{fig:results-cpu-gpu}
        \figggwolegend{results-cpu-gpu}{ylabel = Time per cell update (ns), xlabel = Nodes, log ticks with fixed point}
        }
        \subfloat[Performance per watt ($10^6$ cell updates per second per watt).]{
        \label{fig:results-cpu-gpu-perf-per-watt}
        \figgg{results-cpu-gpu-perf-per-watt}{ylabel = Performance per watt, xlabel = Nodes, log ticks with fixed point, legend pos = outer north east}
        }
        %
        \caption{\revisionadd{A performance comparison of} Astaroth and Pencil Code \revisionadd{(PC-Mahti, PC-Puhti)} in 6th-order stencil computations. A cell update entails the full integration step. The grid sizes \textit{small} (S), \textit{medium} (M), and \textit{large} (L) have been abbreviated for clarity.
        }
\end{figure*}

In our \revisionadd{fifth} test, we compared the strong scaling of Astaroth and PC \revisionadd{(Fig.~\ref{fig:results-cpu-gpu})}. 
\revisionadd{To produce fair benchmarks, PC-Puhti was used with grid dimensions $280^3$, $520^3$, and $1040^3$, which were multiples of the per-node core count and close to the dimensions $256^3$, $512^3$, and $1024^3$ used with Astaroth and PC-Mahti. For clarity, we use the terms \textit{small}, \textit{medium}, and \textit{large} to denote the benchmarks using $256^3$ or $280^3$, $512^3$ or $520^3$, and $1024^3$ or $1040^3$ cells, respectively.}
For decompositions, we chose common-sense processor numbers in the $x$, $y$, and $z$ directions, based on the recommendations of the PC developers and its user manual. However, as the decomposition must be supplied by the user and trying each possible permutation was not feasible within the bounds of this work, our PC benchmarks may not present the best  performance possible.

\revisionadd{On a single node, Astaroth exhibited a $18$--$53\times$ speedup compared to PC. On $16$ nodes, the speedups with Astaroth in the small, medium, and large benchmarks were $4$--$15\times$, $11$--$32\times$, and $20$--$60\times$, respectively.} The scaling efficiency of PC remained relatively high in all tests.

\revisionadd{In our final test (Fig.~\ref{fig:results-cpu-gpu-perf-per-watt}), we measured the energy efficiency of the implementations in terms of cell updates per second per watt. On a single node, we saw $8$--$11\times$ improved energy efficiency with Astroth compared to PC, whereas on $16$ nodes, the energy efficiency improved $2$--$3\times$, $5$--$6\times$, and $9$--$12\times$ in the small, medium, and large benchmarks, respectively.}

\captionsetup[subfigure]{margin=44pt,parskip=5pt,indention=10pt,singlelinecheck=false}


\section{Discussion}
\label{sec:discussion}

The landscape of high-performance computing is evolving: modern compute nodes are heterogeneous, housing multiple specialized accelerators that can provide several times higher throughput in domain-specific tasks. As network bandwidth and latency lag behind operational performance, communication-heavy applications must be carefully tuned to reduce data movement to enable efficient scaling to a large number of nodes. 

In this work, we have addressed the main questions regarding the implementation of
large-scale ISL-based physical simulations on HPC nodes containing multiple GPU devices. By Z-order-based partitioning, and executing computation and communication in parallel, our implementation scaled to $16$ nodes at $\geq 50\%$ efficiency compared to the theoretically achievable performance.

Compared with traditional CPU computations, GPUs can provide competitive throughput \revisionadd{and energy efficiency} in multi-node applications despite additional device-to-host and host-to-device communication latency (\revisionadd{Figs.~\ref{fig:results-cpu-gpu} and~\ref{fig:results-cpu-gpu-perf-per-watt}}). When comparing the operational performance and memory bandwidth available on a single HPC node as used in this work, we expect that GPUs can provide an order of magnitude higher throughput in equivalently optimized data-parallel programs \revisionadd{when the performance is bound by compute. In large-scale computations, the benefits of using GPUs diminish when the network bandwidth becomes the performance limiter. Furthermore on a single node using PC-Puhti, we expected a speedup of roughly $13\times$ with Astaroth based on the available memory bandwidth. Instead, we measured a speedup of $44$--$53\times$, which suggests that the CPU solver was not fully optimized.}

On current hardware in the test cases presented in this work, the benefits of GPUDirect RDMA in high-order stencil computations are modest, as the majority of the messages are above the $32$ KiB threshold (Fig.~\ref{fig:results-isolated}, communication with and without corners). We expect to see more notable benefits from RDMA in larger-scale simulations, where each subgrid contains $\leq 32^3$ cells.

In previous work, we spawned a single process per node, and used CUDA peer-to-peer memory transfers and one-dimensional decomposition for intra-node communication instead of MPI~\cite{vaisala2020}. In this work, we saw no notable performance difference in single-node performance between our previous implementation and our implementation with multiple processes and MPI. The integration time per cell was $0.65$ ns in previous work, in contrast to $0.53$ ns measured in this work. This was expected, as both implementations were compute bound in our tests. Further tests with smaller grid sizes are required to see whether there is a difference when the performance is bound by bandwidth.

Our work has the following limitations. Firstly, we assumed that the bandwidth between any pair of nodes, or devices within a node, is always the same. This holds if the network topology is a fat tree and the devices within a node are fully connected. However, further analysis may be needed to find the optimal decomposition on more complex interconnect topologies. Furthermore, in our model, we assumed that communication can be carried out simultaneously with all neighbors. This is generally not the case on current hardware and the communication cost function should be extended to take this into account for more precise scaling estimates.

\revisionadd{Secondly, our performance model does not account for the communication overhead, or latency, which can take a significant part of the communication time if message sizes are especially small. In this work where large messages were used, we deemed the model to provide sufficiently robust performance estimates.}

\revisionadd{Thirdly}, our implementation is not fully pipelined, which implies lower parallel efficiency in all test cases. This is caused by starting the update of the outer computational segments only after all communication has finished. Scaling efficiency could be improved with a more fine-grained communication scheme, in which the update of an outer segment is started immediately after its data dependencies have been satisfied\revisionadd{~\cite{lappi2021}}.

\revisionadd{Fourthly}, while Morton order indexing provides a good approximation of the optimal decomposition in our test case on an ideal machine, it may not be communication-optimal on practical hardware. Graph-based algorithms for optimizing network communication is a well-studied field~\cite{scotch} and integrating this knowledge into Astaroth is a subject of future work.

As the performance of our implementation was $\geq 50\%$ the theoretical maximum, more aggressive techniques to reduce data movement are required to see significant improvements in scaling efficiency. One such technique, albeit controversial, is real-time data compression. Lossless compression would reduce data movement without loss of precision, however, achieving satisfactory throughput and compression ratio with noisy data may be difficult. Lossy compression typically provides better compression ratios than lossless compression, but may require tuning in accuracy-sensitive applications. A simple form of lossy compression is mixed precision, \revisionadd{supported} for example by GROMACS~\cite{gromacs}, where lower precision is used in calculations where it is known not to result an a catastrophic loss in accuracy. More sophisticated approaches have also been suggested, for example \revisionadd{by Lindstrom}~\cite{lindstrom2014}, where four-fold lossy compression in shock hydrodynamics simulations was reported to result in a relative error of $0.06\%$ after $1000$ timesteps. There has been notable interest towards hardware-accelerated compression~\cite{fowers2015, tavana2019}, as it could provide higher throughput than software implementations. \revisionadd{Preliminary studies have also been conducted on using machine learning to reconstruct physically accurate turbulent flows from low-resolution data~\cite{guemes2021, kim2021}.}

Whether the disproportional growth of operational, memory, and network performance continues is an open question. At the current rate, the performance of all implementations will eventually become bound by data movement. However, the off-chip memory bottleneck has been notably alleviated, but not eliminated, with the introduction of 3D-stacked CMOS technology~\cite{jacob2009}. Technologies based on silicon photonics are expected to bring similar improvements to network performance~\cite{rumley2015}. Moreover, the physical limits for transistor densities in silicon-based microprocessors are expected to be encountered around 2025~\cite{waldrop2016, Shalf2020}. As the die size cannot be increased indefinitely to fit more transistors due to lower manufacturing efficiency and issues with heat dissipation~\cite{tavana2019}, new technologies are needed to see continued improvements in operational performance. Several technologies have been suggested, such as graphene-based microprocessors and integrating multiple GPU modules on a single package~\cite{arunkumar2017}.

Until physical manufacturing limits can be overcome, we expect to see increased use of domain-specific accelerators, such as GPUs, TPUs, and ASICs, due to their ability to provide higher throughput and energy efficiency with existing manufacturing technologies than general-purpose processors. In this light, we believe that data movement will continue to be a major challenge in large-scale computations.

\section{Conclusion}
\label{sec:conclusion}

Computing centers have started to incorporate specialized accelerators to HPC nodes to improve
throughput and power efficiency in domain-specific tasks. This has further increased the 
gap between computational performance and the rate at which data can be transferred
via the network. To enable efficient scaling to multiple nodes in communication-heavy
applications, minimizing inter- and intra-node communication is of utmost priority.

In this work, we presented an analysis of the scaling of iterative stencil loops and applied
established techniques to implement a scalable multi-GPU communication scheme, which we
evaluated in high-order MHD simulations. In our benchmarks, we saw that the per-node performance
improvement from GPUs outweighs the added device-to-host and host-to-device communication latencies,
and that strong scaling to at least $64$ GPUs is possible at high efficiency with sufficiently large problem sizes.
Because inter-node bandwidth is an expensive resource, improving intra-node data locality and reducing overall data movement, even at the cost of redundant computations, is likely required to see more efficient scaling in communication-bound problems.

\section*{Acknowledgements}

This study is a product of Astaroth collaboration. The authors thank Fredrik Robertsén (CSC) for valuable discussion and technical support. J.P., M.J.K., and M.R.\ acknowledge the support of the Academy of Finland
ReSoLVE Centre of Excellence (grant number 307411).
This project has received funding from the European Research Council (ERC)
under the European Union's Horizon 2020 research and innovation
programme (Project UniSDyn, grant agreement n:o 818665).
M.V. acknowledges funding support for Theory within ASIAA from Academia Sinica.

\bibliographystyle{elsarticle-num}
\bibliography{references.bib}

\clearpage
\newpage
\section*{Appendix A: Decomposition}
\renewcommand{\thetable}{A.\arabic{table}}
\setcounter{table}{0}

Solutions to Eq.~\ref{eq:decomposition-ip} for typical problem sizes, in which the stencil contains all points within its radius and the boundaries are periodic. We evaluated each valid decomposition using a brute-force search. Only one decomposition is listed if there is more than one solution. The optimal inter-node level decompositions $P$ are shown in Table~\ref{tab:qmin-inter}. Intra-node level decompositions $P'$ are shown in Table~\ref{tab:qmin-intra}.

\begin{table}[t]
    \caption{Values of $P$ that solve Eq.~\ref{eq:decomposition-ip} for typical three-dimensional problem sizes.}
    \label{tab:qmin-inter}
    \vfill
    \subtbl{qmin-a}{$N = (512, 512, 512)$.}
    \hfill
    \subtbl{qmin-b}{$N = (1024, 512, 512)$.}
    \hfill
    \subtbl{qmin-c}{$N = (1024, 1024, 512)$.}
\end{table}

\begin{table}[t]
    \caption{Values of $P'$ that solve Eq.~\ref{eq:decomposition-ip} adapted for intra-node communication for typical three-dimensional problem sizes.}
    \label{tab:qmin-intra}
    \vfill
    \subtbl{qmin-intra-a}{$N' = (512, 512, 512)$.}
    \hfill
    \subtbl{qmin-intra-b}{$N' = (1024, 512, 512)$.}
    \hfill
    \subtbl{qmin-intra-c}{$N' = (1024, 1024, 512)$.}
\end{table}

\newpage
\section*{Appendix B: MHD equations}
\renewcommand{\thetable}{B.\arabic{table}}
\setcounter{table}{0}
\renewcommand{\theequation}{B.\arabic{equation}}
\setcounter{equation}{0}

The basic physical quantities updated during each 
Runge-Kutta timestep are shown in Table~\ref{table:fields}. We use the 
standard non--ideal MHD equations in non-conservative form: 
\begin{align}
\begin{split}
\label{eq:continuity}
\frac{D \ln\rho}{D t} ={}& - \nabla \cdot \mathbf{u} \ ;
\end{split}\\
\begin{split}\label{eq:navierstokes}
\frac{D \mathbf{u}}{D t} ={}& -c_s^2 \nabla \biggl(\frac{s}{c_{p}} + \ln\rho \biggr) + \frac{\mathbf{j} \times \mathbf{B}}{\rho} \\
                            & +\nu \bigg[ \nabla^2\mathbf{u} + \frac{1}{3}\nabla(\nabla \cdot \mathbf{u}) + 2 \mathbf{S} \cdot \nabla \ln \rho \bigg]\\
                            & + \zeta \nabla(\nabla \cdot \mathbf{u}) \ .
\end{split} \\
\begin{split}\label{eq:entropy}
\rho T \frac{Ds}{Dt} ={}& \mathcal{H} - \mathcal{C} + \nabla \cdot (K \nabla T) + \eta \mu_0 \mathbf{j}^2 \\
                        &+ 2 \rho \nu \mathbf{S} \otimes \mathbf{S} + \zeta \rho(\nabla \cdot \mathbf{u})^2 \ .
\end{split}\\
\begin{split}
\label{eq:magnetic}
\frac{\partial \mathbf{A}}{\partial t} ={}& \mathbf{u} \times \mathbf{B} 
+ \eta \nabla^2 \mathbf{A} \ .
\end{split}
\end{align}
See Table~\ref{table:symbols} for symbol explanations.
We refer the reader to~\cite{brandenburg2003} for a more
detailed discussion on the physical system and related computational aspects.
 

\begin{table}
  \begin{center}
\caption{Basic physical quantities solved for.}
    \label{table:fields}
    \begin{filecontents*}{fields.dat}
%% LaTeX2e file `fields.dat'
%% generated by the `filecontents' environment
%% from source `main' on 2022/05/10.
%%
Field, Symbol
Logarithmic density, $\ln \rho$
Velocity, $\mathbf{u}$
Specific entropy, $s$
Magnetic vector potential, $\mathbf{A}$
\end{filecontents*}
\pgfplotstabletypeset[
      multicolumn names, 
      col sep=comma, 
      display columns/0/.style={
		column name=Field,
		string type,
        column type = l
        },
      display columns/1/.style={
		column name=Symbol,
		string type,
        column type = l
        },
      every head row/.style={
		before row={\toprule}, 
		after row={\midrule} 
			},
		every last row/.style={after row=\bottomrule}, 
    ]{fields.dat} 
  \end{center}
\end{table}

\begin{table}
  \begin{center}
    \caption{Other symbols.}
    \label{table:symbols}
    \begin{filecontents*}{symbols.dat}
%% LaTeX2e file `symbols.dat'
%% generated by the `filecontents' environment
%% from source `main' on 2022/05/10.
%%
Field, Symbol
Laplace operator, $\nabla^2$
Curl operator, $\nabla \times$
Advective derivative, $D / Dt = \partial / \partial t + \mathbf{u} \cdot \nabla $ %MJK
%Magnetic flux density, $\mathbf{B}$
Magnetic field, $\mathbf{B}=\nabla \times \mathbf{A}$
Magnetic diffusivity, $\eta$
Magnetic vacuum permeability, $\mu_0$
Electric current density, $\mathbf{j}=\mu_0^{-1} \nabla \times \mathbf{B}$
Traceless rate-of-shear tensor, $\mathbf{S}$
Specific heat capacity at constant pressure, $c_p$
Specific heat capacity at constant volume, $c_v$
Kinematic viscosity, $\nu$
Bulk viscosity, $\zeta$
Adiabatic speed of sound, $c_s$
Adiabatic index, $\gamma$
Explicit heating term, $\mathcal{H}$
Explicit cooling term, $\mathcal{C}$
Radiative thermal conductivity, $K$
Temperature, $T$
%MR: not used
%Thermal diffusivity, $\chi$
\end{filecontents*}
\pgfplotstabletypeset[
      multicolumn names, 
      col sep=comma, 
      display columns/0/.style={
		column name=Description,
		string type,
        column type = l
        },
      display columns/1/.style={
		column name=Symbol,
		string type,
        column type = l
        },
      every head row/.style={
		before row={\toprule}, 
		after row={\midrule} 
			},
		every last row/.style={after row=\bottomrule}, 
    ]{symbols.dat} 
  \end{center}
\end{table}

\end{document}

%% file: commands.tex
\newcommand{\subtbl}[2]{
	\begin{subtable}[h]{0.30\textwidth}
		\centering
		\caption{#2}
		\label{tab:#1}
		\pgfplotstabletypeset [
			col sep = comma,
			every head row/.style = {
				before row = \toprule,
				after row  = \midrule,
			},
			every last row/.style = {
				after row = \bottomrule
			},
		]{data/#1.csv}
	\end{subtable}
}

\newcommand{\plotfile}[1]{
    \pgfplotstableread{#1}{\table}
    \pgfplotstablegetcolsof{#1}
    \pgfmathtruncatemacro\numberofcols{\pgfplotsretval-1}
    \pgfplotsinvokeforeach{1,...,\numberofcols}{
        \pgfplotstablegetcolumnnamebyindex{##1}\of{\table}\to{\colname}
        \addplot table [y index=##1] {#1}; 
        \addlegendentryexpanded{\colname}
    }
}

\pgfplotsset{
  log x ticks with fixed point/.style={
      xticklabel={
        \pgfkeys{/pgf/fpu=true}
        \pgfmathparse{2^(\tick)}%
        \pgfmathprintnumber[fixed relative, precision=3]{\pgfmathresult}
        \pgfkeys{/pgf/fpu=false}
      }
  },
  log y ticks with fixed point/.style={
      yticklabel={
        \pgfkeys{/pgf/fpu=true}
        \pgfmathparse{2^(\tick)}%
        \pgfmathprintnumber[fixed relative, precision=3]{\pgfmathresult}
        \pgfkeys{/pgf/fpu=false}
      }
  }
}

\newcommand{\fig}[3]{
	\begin{figure}
			\centering
			\begin{tikzpicture}[draw = black, text = black]
					\pgfplotsset{width = 0.85\columnwidth, height = 0.64\columnwidth}
					\begin{axis} [
					        #3,
							xmode = log,
							log basis x = 2,
						    log x ticks with fixed point,
						    mark list fill=.!50!white,
							ymode = log,
							grid = major,
							legend cell align = left,
							draw = black,
                            text = black,
                            cycle list name = mark list
					]
					\plotfile{data/#1.csv}
					\end{axis}
			\end{tikzpicture}
			\caption{#2}
			\label{fig:#1}
	\end{figure}
}

\newcommand{\figg}[2]{
			\begin{tikzpicture}[draw = black, text = black]
					\pgfplotsset{width = 0.85\columnwidth, height = 0.64\columnwidth}
 
					\begin{axis} [
					        #2,
							xmode = log,
							log basis x = 2,
						    log x ticks with fixed point,
						    mark list fill=.!50!white,
							ymode = log,
							grid = major,
							legend cell align = left,
							draw = black,
                            text = black,
                            cycle list name = mark list
					]
					\plotfile{data/#1.csv}
					\end{axis}
			\end{tikzpicture}
}

\newcommand{\figgg}[2]{
			\begin{tikzpicture}[draw = black, text = black]
					\pgfplotsset{width = 0.85\columnwidth, height = 0.64\columnwidth}
 
					\begin{axis} [
							xmode = log,
							log basis x = 2,
						    log x ticks with fixed point,
							ymode = log,
							grid = major,
							legend cell align = left,
							draw = black,
                            text = black,
                            mark list fill=.!50!white,
                            cycle multi list = {
                                linestyles\nextlist
                                [3 of]mark list
                            },
                            #2
					]
					\plotfile{data/#1.csv}
					\end{axis}
			\end{tikzpicture}
}

\newcommand{\figggwolegend}[2]{
			\begin{tikzpicture}[draw = black, text = black]
					\pgfplotsset{width = 0.85\columnwidth, height = 0.64\columnwidth}
 
					\begin{axis} [
							xmode = log,
							log basis x = 2,
						    log x ticks with fixed point,
							ymode = log,
							grid = major,
							legend cell align = left,
							draw = black,
                            text = black,
                            mark list fill=.!50!white,
                            cycle multi list = {
                                linestyles\nextlist
                                [3 of]mark list
                            },
                            #2
					]
					\plotfile{data/#1.csv}
					\legend{};
					\end{axis}
			\end{tikzpicture}
}

%% file: tikz/fig-stencil.tex
\newcommand{\stencila}[3] {
    \foreach \i in {1, ..., #3} {
        \draw[color = black, fill = gray, shift = {(#1 + \i, #2)}] rectangle(1, 1);
        \draw[color = black, fill = gray, shift = {(#1 - \i, #2)}] rectangle(1, 1);
        \draw[color = black, fill = gray, shift = {(#1, #2 + \i)}] rectangle(1, 1);
        \draw[color = black, fill = gray, shift = {(#1, #2 - \i)}] rectangle(1, 1);
    }
}

\newcommand{\stencilb}[3] {
    \foreach \i in {1, ..., #3} {
        \draw[color = black, fill = gray, shift = {(#1 + \i, #2 + \i)}] rectangle(1, 1);
        \draw[color = black, fill = gray, shift = {(#1 + \i, #2 - \i)}] rectangle(1, 1);
        \draw[color = black, fill = gray, shift = {(#1 - \i, #2 + \i)}] rectangle(1, 1);
        \draw[color = black, fill = gray, shift = {(#1 - \i, #2 - \i)}] rectangle(1, 1);
    }
    \draw[color = black, fill = white, shift = {(#1, #2)}] rectangle(1, 1);
}

\newcommand{\stencilc}[3] {
    \foreach \i in {1, ..., #3} {
        \draw[color = black, fill = gray, shift = {(#1 + \i, #2 + \i)}] rectangle(1, 1);
        \draw[color = black, fill = gray, shift = {(#1 + \i, #2 - \i)}] rectangle(1, 1);
        \draw[color = black, fill = gray, shift = {(#1 - \i, #2 + \i)}] rectangle(1, 1);
        \draw[color = black, fill = gray, shift = {(#1 - \i, #2 - \i)}] rectangle(1, 1);
        
        \draw[color = black, fill = gray, shift = {(#1 + \i, #2)}] rectangle(1, 1);
        \draw[color = black, fill = gray, shift = {(#1 - \i, #2)}] rectangle(1, 1);
        \draw[color = black, fill = gray, shift = {(#1, #2 + \i)}] rectangle(1, 1);
        \draw[color = black, fill = gray, shift = {(#1, #2 - \i)}] rectangle(1, 1);
    }
}

\newcommand{\stencild}[3] {
    \foreach \j in {1, ..., #3} {
        \foreach \i in {1, ..., #3} {
            \draw[color = black, fill = gray, shift = {(#1 + \i, #2 + \j)}] rectangle(1, 1);
            \draw[color = black, fill = gray, shift = {(#1 + \i, #2 - \j)}] rectangle(1, 1);
            \draw[color = black, fill = gray, shift = {(#1 - \i, #2 + \j)}] rectangle(1, 1);
            \draw[color = black, fill = gray, shift = {(#1 - \i, #2 - \j)}] rectangle(1, 1);
            
            \draw[color = black, fill = gray, shift = {(#1 + \i, #2)}] rectangle(1, 1);
            \draw[color = black, fill = gray, shift = {(#1 - \i, #2)}] rectangle(1, 1);
            \draw[color = black, fill = gray, shift = {(#1, #2 + \j)}] rectangle(1, 1);
            \draw[color = black, fill = gray, shift = {(#1, #2 - \j)}] rectangle(1, 1);
        }
    }
}

\begin{tikzpicture}[scale=0.30, draw = black, text = black]
    \pgfmathsetmacro{\n}{4.0}
    \pgfmathsetmacro{\r}{0.1 * \n}
    \pgfmathsetmacro{\m}{\n + 2.0 * \r}

    \pgfmathsetmacro{\ns}{\n / 2.0}
    \pgfmathsetmacro{\ms}{\ns + 2.0 * \r}

    \pgfmathsetmacro{\es}{3}
    \pgfmathsetmacro{\ds}{\r / \es}

    \stencila{0}{0}{2}
    \stencilb{7}{0}{2}
    \stencilc{14}{0}{2}
    \stencild{21}{0}{2}

    \draw[<->, fill = white, shift = {(25, 1)}] (0, 0) -- (0, 2) node[midway, right] {$r$};
\end{tikzpicture}

%% file: tikz/fig-grid.tex
  \begin{tikzpicture}[scale=0.9, draw = black, text = black]

        \pgfmathsetmacro{\n}{4.0}
        \pgfmathsetmacro{\r}{0.1 * \n}
        \pgfmathsetmacro{\m}{\n + 2.0 * \r}
        
        \pgfmathsetmacro{\ns}{\n / 2.0}
        \pgfmathsetmacro{\ms}{\ns + 2.0 * \r}
        
        \pgfmathsetmacro{\es}{3}
        \pgfmathsetmacro{\ds}{\r / \es}
        
        
        
        \draw[fill=white, shift = {(\r, \r)}] rectangle(\ns, \ns);
        \draw[fill=white, shift = {(\r + \ns, \r)}] rectangle(\ns, \ns);
        \draw[fill=white, shift = {(\r + \ns, \r + \ns)}] rectangle(\ns, \ns);
        \draw[fill=white, shift = {(\r, \r + \ns)}] rectangle(\ns, \ns);
        
        \draw[dashed, shift = {(0, 0)}] rectangle (\ms, \ms);
        \fill[gray, opacity = 0.2, shift = {(0, 0)}] rectangle (\ms, \ms);
        \draw[fill = white, shift = {(\r, \r)}] rectangle(\ns, \ns);
        
        
        \draw[dotted, shift = {(0, 0)}] rectangle (\ms, \ms);
        \draw[dotted, shift = {(\ns, 0)}] rectangle (\ms, \ms);
        \draw[dotted, shift = {(\ns, \ns)}] rectangle (\ms, \ms);
        \draw[dotted, shift = {(0, \ns)}] rectangle (\ms, \ms);
        
        \draw[<->, shift = {(\m + \r, \r)}] (0, \n) -- (0, 0) node[midway, right] {$n_1$};
        \draw[<->, shift = {(\m + \r + 0.75, 0)}] (0, \m) -- (0, 0) node[midway, right] {$m_1$};
        
        \draw[<->, shift = {(-\r, \r)}] (0, \ns) -- (0, 0) node[midway, left] {$n_1'$};
        \draw[<->, shift = {(-\r - 0.75, 0)}] (0, \ms) -- (0, 0) node[midway, left] {$m_1'$};
        
        \draw[<->, shift = {(-\r, \m)}] (0, 0) -- (0, -\r) node[midway, left] {$r$};

            
        
        
        \draw[shift = {(0, \m + 2.0*\r - 0.0*\r)}] rectangle(\r, \r) node[shift = {(0.5*\r, 0)}, midway, right] {Computational subdomain $N'$};
        \draw[fill = gray, fill opacity = 0.2, shift = {(0,  \m + 2.0*\r - 1.0*\r)}] rectangle(\r, \r) node[opacity = 1.0, shift = {(0.5*\r, 0)}, midway, right] {Halo};
  \end{tikzpicture}

%% file: tikz/fig-downhill.tex
\begin{tikzpicture}
  \pgfplotsset{                                                                  
    width = \columnwidth, height = 0.6\columnwidth                                                           
  }                                                                              
  \begin{axis}[                                                                  
      ylabel = Running time,                                                       
      xlabel = Number of processing elements,                                              
      xticklabels = {,,},                                                        
      yticklabels = {,,},                                                        
      legend cell align = left,                                                  
      legend pos = north east,                                                   
      axis lines = left, 
      ticks = none,                                                              
      xmin = 1,                                                                  
      xmax  = 12,                                                                
      ymin = 0,                                                                  
      ymax = 8,                 
      draw = black,
      text = black,
    ]                                                       

    \draw (0, 7) -- (6, 2) node [midway, above right] {$\tau_W$};
    \draw (6, 2) -- (11, 0.75) node [midway, above right] {$\tau_Q$};                
                                                                                 
    \addplot [dotted, domain = 6:8] {-x + 8};                                    
    \addplot [dotted, domain = 1:6] {-x/4 + 3.5};                                
                                      
    \draw [dashed ](6, 0) -- (6, 5.25);                                           
    \draw [->] (6, 1) -- (5, 1) node [align = center, left] {Compute\\bound};
    \draw [->] (6, 4) -- (7, 4) node [align = center, right] {Communication\\bound};
                                                                                 
                                                                                 
  \end{axis}                                                                     
\end{tikzpicture}

%% file: tikz/fig-subgrid.tex
  \begin{tikzpicture}[scale=0.9, draw = black, text = black]

        \pgfmathsetmacro{\n}{4.0}
        \pgfmathsetmacro{\r}{0.1 * \n}
        \pgfmathsetmacro{\m}{\n + 2.0 * \r}
        
        \pgfmathsetmacro{\ns}{\n / 2.0}
        \pgfmathsetmacro{\ms}{\ns + 2.0 * \r}
        
        \pgfmathsetmacro{\es}{3}
        \pgfmathsetmacro{\ds}{\r / \es}
        
        
        
        \draw[fill=white, shift = {(\r, \r)}] rectangle(\ns, \ns);
        \draw[fill=white, shift = {(\r + \ns, \r)}] rectangle(\ns, \ns);
        \draw[fill=white, shift = {(\r + \ns, \r + \ns)}] rectangle(\ns, \ns);
        \draw[fill=white, shift = {(\r, \r + \ns)}] rectangle(\ns, \ns);
        
        \draw[dashed, shift = {(0, 0)}] rectangle (\ms, \ms);
        \draw[fill = white, shift = {(\r, \r)}] rectangle(\ns, \ns);
        
        \fill[gray, opacity = 0.2, shift = {(\r, \r)}] rectangle (\ms - \r, \ms - \r);
        
        \draw[dotted, shift = {(0, 0)}] rectangle (\ms, \ms);
        \draw[dotted, shift = {(\ns, 0)}] rectangle (\ms, \ms);
        \draw[dotted, shift = {(\ns, \ns)}] rectangle (\ms, \ms);
        \draw[dotted, shift = {(0, \ns)}] rectangle (\ms, \ms);
        
        \draw[<->, shift = {(\m + \r, \r)}] (0, \n) -- (0, 0) node[midway, right] {$n_1'$};
        \draw[<->, shift = {(\m + \r + 0.75, 0)}] (0, \m) -- (0, 0) node[midway, right] {$m_1'$};
        
        \draw[<->, shift = {(-\r, \r)}] (0, \ns) -- (0, 0) node[midway, left] {$n_1''$};
        \draw[<->, shift = {(-\r - 0.75, 0)}] (0, \ms) -- (0, 0) node[midway, left] {$m_1''$};
        
        \draw[<->, shift = {(-\r, \m)}] (0, 0) -- (0, -\r) node[midway, left] {$r$};

            
        
        
        \draw[shift = {(0, \m + 2.0*\r - 0.0*\r)}] rectangle(\r, \r) node[shift = {(0.5*\r, 0)}, midway, right] {Computational subdomain $N''$};
        \draw[fill = gray, fill opacity = 0.2, shift = {(0,  \m + 2.0*\r - 1.0*\r)}] rectangle(\r, \r) node[opacity = 1.0, shift = {(0.5*\r, 0)}, midway, right] {Local subgrid $L''$};
  \end{tikzpicture}

%% file: tikz/fig-sfc.tex
\newcommand\scannode[2] {
    \begin{scope}[shift = {(#1, #2)}]
        \foreach \i in {1, ..., 3} {
            \draw[dotted] (\i, 0) -- (\i, 1);
        }
        
        \draw[white, thick] (0, 0) -- (4, 0) -- (4, 1) -- (0, 1) -- cycle;
        \draw[dashed] (0, 0) -- (4, 0) -- (4, 1) -- (0, 1) -- cycle;
    \end{scope}
}
\newcommand\mortonnode[2] {
    \begin{scope}[shift = {(#1, #2)}]
        \draw[dotted] (1, 0) -- (1, 2);
        \draw[dotted] (0, 1) -- (2, 1);
        
        \draw[white, thick] (0, 0) -- (2, 0) -- (2, 2) -- (0, 2) -- cycle;
        \draw[dashed] (0, 0) -- (2, 0) -- (2, 2) -- (0, 2) -- cycle;
    \end{scope}
}
\newcommand\mortonx[1] {
    {mod((#1), 2) + 2 * mod(floor((#1)/4), 2) + 4 * mod(floor((#1)/16), 2) + 8 * mod(floor((#1)/64), 2)}
}
\newcommand\mortony[1] {
    {mod((floor((#1)/2)), 2) + 2 * mod(floor((floor((#1)/2))/4), 2) + 4 * mod(floor((floor((#1)/2))/16), 2) + 8 * mod(floor((floor((#1)/2))/64), 2)}
}

\newcommand\rowwisescan {
  \begin{tikzpicture} [scale = 0.48, draw = black, text = black]
    \begin{scope}[shift = {(0, 0)}]
        \foreach \j in {1, ..., 8} {
            \scannode{0}{\j}
            \scannode{4}{\j}
        }
        \foreach \j in {1, ..., 8} {
            \draw[thick] (0.5, \j + 0.5) -- (7.5, \j + 0.5);
        }
        \foreach \j in {2, ..., 8} {
            \draw[thick] (0.5, \j + 0.5) -- (7.5, \j - 0.5);
        }
    \end{scope}
    \end{tikzpicture}
}

\newcommand\mortonorder {
    \begin{tikzpicture} [scale = 0.48, draw = black, text = black]
    \begin{scope}[shift = {(0, 0)}]
        \foreach \j in {0, ..., 3} {
            \mortonnode{0}{2 * \j}
            \mortonnode{2}{2 * \j}
            \mortonnode{4}{2 * \j}
            \mortonnode{6}{2 * \j}

            \foreach \i in {0, ..., 3} {
                \begin{scope}[shift = {(2 * \i + 0.5, 2 * \j + 0.5)}]
                \end{scope}
            }
        \foreach \i in {1, ..., 63} {
            \def\prevx{\mortonx{\i-1}}
            \def\prevy{\mortony{\i-1}}
            \def\currx{\mortonx{\i}}
            \def\curry{\mortony{\i}}

            \begin{scope}[shift = {(0.5, 0.5)}]
            \draw[thick] (\prevx, \prevy) -- (\currx, \curry);
            \end{scope}
        }
        }
    \end{scope}
  \end{tikzpicture}
}
\newcommand\hilbertcurve{
    \begin{tikzpicture} [scale = 0.32]
    \begin{scope}[shift = {(0, 0)}]
        \foreach \j in {0, ..., 3} {
            \mortonnode{0}{2 * \j}
            \mortonnode{2}{2 * \j}
            \mortonnode{4}{2 * \j}
            \mortonnode{6}{2 * \j}

            \foreach \i in {0, ..., 3} {
                \begin{scope}[shift = {(2 * \i + 0.5, 2 * \j + 0.5)}]
                \end{scope}
            }
        }
        \pgfdeclarelindenmayersystem{Hilbert curve}{
          \rule{L -> +RF-LFL-FR+}
          \rule{R -> -LF+RFR+FL-}}
          \begin{scope}[shift = {(0.5, 0.5)}]
            \draw[thick]
            [l-system={Hilbert curve, axiom=L, order=3, step=28.16pt, angle=90}] lindenmayer system;
          \end{scope}
    \end{scope}
  \end{tikzpicture}
}

%% file: tikz/fig-segments.tex
\newcommand\cube[3] {                                                       
  (0, 0, #3) -- (#1, 0, #3) -- (#1, #2, #3) -- (0, #2, #3) -- cycle         
  (0, #2, 0) -- (#1, #2, 0) -- (#1, #2, #3) -- (0, #2, #3) -- cycle         
  (#1, 0, 0) -- (#1, #2, 0) -- (#1, #2, #3) -- (#1, 0, #3) -- cycle             
}

\begin{tikzpicture}[x = {(-0.866cm, -0.5cm)}, z = {(0.866cm,-0.5cm)}, y = {(0cm,1cm)}, scale=2.0, draw = black, text = black]
  \pgfmathsetmacro{\n} {1}
  \pgfmathsetmacro{\r} {3/16}
  \pgfmathsetmacro{\m} {\n + 2 * \r}
  \pgfmathsetmacro{\sep} {0.5 * \r}

  \draw[fill = white, shift = {(0, 0, 0)}] \cube{\r}{\r}{\r};
  \draw[fill = white, shift = {(\r + \sep, 0, 0)}] \cube{\n}{\r}{\r};
  \draw[fill = white, shift = {(\r + \sep + \n + \sep, 0, 0)}] \cube{\r}{\r}{\r};

  \draw[fill = white, shift = {(0, \r + \sep, 0)}] \cube{\r}{\n}{\r};
  \draw[fill = white, shift = {(\r + \sep, \r + \sep, 0)}] \cube{\n}{\n}{\r};
  \draw[fill = white, shift = {(\r + \sep + \n + \sep, \r + \sep, 0)}] \cube{\r}{\n}{\r};

  \draw[fill = white, shift = {(0, \r + \sep + \n + \sep, 0)}] \cube{\r}{\r}{\r};
  \draw[fill = white, shift = {(\r + \sep, \r + \sep + \n + \sep, 0)}] \cube{\n}{\r}{\r};
  \draw[fill = white, shift = {(\r + \sep + \n + \sep, \r + \sep + \n + \sep, 0)}] \cube{\r}{\r}{\r};

  \draw[fill = white, shift = {(0, 0, \r + \sep)}] \cube{\r}{\r}{\n};
  \draw[fill = white, shift = {(\r + \sep, 0, \r + \sep)}] \cube{\n}{\r}{\n};
  \draw[fill = white, shift = {(\r + \sep + \n + \sep, 0, \r + \sep)}] \cube{\r}{\r}{\n};

  \draw[fill = white, shift = {(0, \r + \sep, \r + \sep)}] \cube{\r}{\n}{\n};
  \draw[fill = white, shift = {(\r + \sep, \r + \sep, \r + \sep)}] \cube{\n}{\n}{\n};
  \draw[fill = white, shift = {(\r + \sep + \n + \sep, \r + \sep, \r + \sep)}] \cube{\r}{\n}{\n};

  \draw[fill = white, shift = {(0, \r + \sep + \n + \sep, \r + \sep)}] \cube{\r}{\r}{\n};
  \draw[fill = white, shift = {(\r + \sep, \r + \sep + \n + \sep, \r + \sep)}] \cube{\n}{\r}{\n};
  \draw[fill = white, shift = {(\r + \sep + \n + \sep, \r + \sep + \n + \sep, \r + \sep)}] \cube{\r}{\r}{\n};

  \draw[fill = white, shift = {(0, 0, \r + \sep + \n + \sep)}] \cube{\r}{\r}{\r};
  \draw[fill = white, shift = {(\r + \sep, 0, \r + \sep + \n + \sep)}] \cube{\n}{\r}{\r};
  \draw[fill = white, shift = {(\r + \sep + \n + \sep, 0, \r + \sep + \n + \sep)}] \cube{\r}{\r}{\r};

  \draw[fill = white, shift = {(0, \r + \sep, \r + \sep + \n + \sep)}] \cube{\r}{\n}{\r};
  \draw[fill = white, shift = {(\r + \sep, \r + \sep, \r + \sep + \n + \sep)}] \cube{\n}{\n}{\r};
  \draw[fill = white, shift = {(\r + \sep + \n + \sep, \r + \sep, \r + \sep + \n + \sep)}] \cube{\r}{\n}{\r};

  \draw[fill = white, shift = {(0, \r + \sep + \n + \sep, \r + \sep + \n + \sep)}] \cube{\r}{\r}{\r};
  \draw[fill = white, shift = {(\r + \sep, \r + \sep + \n + \sep, \r + \sep + \n + \sep)}] \cube{\n}{\r}{\r};
  \draw[fill = white, shift = {(\r + \sep + \n + \sep, \r + \sep + \n + \sep, \r + \sep + \n + \sep)}] \cube{\r}{\r}{\r};
\end{tikzpicture}

%% file: tikz/fig-iteration.tex
\begin{tikzpicture}[xscale = 1.0, draw = black, text = black]
    
        \def\wpack{1.0}
        \def\wupack{1.1}
        \def\wexchange{5.6}
        \def\winner{\wexchange - \wpack/2 - \wupack/2}
        \def\wouter{1.5}

        \draw (0, -1) rectangle ++(\wpack, 0.5) node[midway] {pack};
        \draw (\wpack, 0) rectangle ++(\winner, 0.5) node[midway] {inner segment update};
        \draw (0.5 * \wpack, -1.5) rectangle ++(\wexchange, 0.5) node[midway] {halo exchange};
        \draw (0.5 * \wpack + \wexchange - \wupack/2, -2) rectangle ++(\wupack, 0.5) node[midway] {unpack};
        \draw (0.5 * \wpack + \wexchange + \wupack/2, -2) rectangle ++(\wouter, 0.5) node[midway] {outer seg.};
        
        \draw [thin, dashed] (0, -2) -- ++(0, 2.5);
        \draw [thin, dashed] (0.5 * \wpack + \wexchange + \wupack/2 + \wouter, -2) -- ++(0, 2.5);
        
        \draw [->] (0, 1.25) -- (2, 1.25) node [midway, above] {time};

\end{tikzpicture}

%% file: tikz/fig-stencil-fd.tex
\newcommand{\stencil}[3] {
\foreach \i in {1, ..., #3} {
            \draw[color = black, fill = gray, shift = {(#1 + \i, #2 + \i)}] rectangle(1, 1);
            \draw[color = black, fill = gray, shift = {(#1 + \i, #2 - \i)}] rectangle(1, 1);
            \draw[color = black, fill = gray, shift = {(#1 - \i, #2 + \i)}] rectangle(1, 1);
            \draw[color = black, fill = gray, shift = {(#1 - \i, #2 - \i)}] rectangle(1, 1);
            
            \draw[color = black, fill = gray, shift = {(#1 + \i, #2)}] rectangle(1, 1);
            \draw[color = black, fill = gray, shift = {(#1 - \i, #2)}] rectangle(1, 1);
            \draw[color = black, fill = gray, shift = {(#1, #2 + \i)}] rectangle(1, 1);
            \draw[color = black, fill = gray, shift = {(#1, #2 - \i)}] rectangle(1, 1);
        }
}

\begin{tikzpicture}[scale=0.26, draw = black, text = black]
    \pgfmathsetmacro{\n}{4.0}
    \pgfmathsetmacro{\r}{0.1 * \n}
    \pgfmathsetmacro{\m}{\n + 2.0 * \r}

    \pgfmathsetmacro{\ns}{\n / 2.0}
    \pgfmathsetmacro{\ms}{\ns + 2.0 * \r}

    \pgfmathsetmacro{\es}{3}
    \pgfmathsetmacro{\ds}{\r / \es}

    \stencil{0}{0}{1}
    \stencil{5}{0}{2}
    \stencil{12}{0}{3}
    \stencil{21}{0}{4}

    \draw[<->, fill = white, shift = {(27, 1)}] (0, 0) -- (0, 4) node[midway, right] {$r$};
\end{tikzpicture}